\documentclass[%
 aip,
 jmp,%
 amsmath,amssymb,
 reprint,%
]{revtex4-1}
\usepackage{graphicx}% Include figure files
\usepackage{dcolumn}% Align table columns on decimal point
\usepackage{bm}% bold math
\usepackage{color}

\begin{document}

\title
{Oxygen modulated quantum conductance for ultra-thin HfO$_2$-based memristive switching devices}
\author{Xiaoliang Zhong}
\email{zhongx@anl.gov}
\affiliation{Materials Science Division, Argonne National Laboratory, Lemont, Illinois 60439, USA}

\author{Ivan Rungger}
\affiliation{Materials Division, National Physical Laboratory, Teddington, TW11 0LW, United Kingdom}

\author{Peter Zapol}
\affiliation{Materials Science Division, Argonne National Laboratory, Lemont, Illinois 60439, USA}

\author{Olle Heinonen}
\email{heinonen@anl.gov}
\affiliation{Materials Science Division, Argonne National Laboratory, Lemont, Illinois 60439, USA}
\affiliation{Northwestern-Argonne Institute for Science and Engineering, Northwestern University, 2145 Sheridan Rd., Evanston, Illinois 60208, USA}

%\begin{document}

\begin{abstract}
Memristive switching devices, candidates for resistive random access memory technology, have been shown to switch off through a progression of states with quantized conductance and subsequent non-integer conductance (in terms of conductance quantum $G_0$). We have performed calculations based on density functional theory to model the switching process for a Pt-HfO$_2$-Pt structure, involving the movement of one or two oxygen atoms. Oxygen atoms moving within a conductive oxygen vacancy filament act as tunneling barriers, and partition the filament into weakly coupled quantum wells. We show that the low-bias conductance decreases exponentially when one oxygen atom moves away from interface. Our results demonstrate the high sensitivity of the device conductance to the position of oxygen atoms.  

%Keywords: HfO$_2$, memristors, switching mechanism, quantized conductance, oxygen migration 
\end{abstract}
\maketitle

Memristive switching devices are currently intensively investigated as potential candidates for non-volatile resistive random access memories because of their scalablility, low-power and high-speed features\cite{yang2013memristive}. The architecture of a typical device consists of a thin film of resistive switching material sandwiched between two electrodes. Recently, quantized conductance has been observed for various resistive switching materials, e.g, AgI\cite{0957-4484-23-14-145703}, Ag$_2$S\cite{terabe2005quantized, wagenaar2012observing}, and a range of transition  metal oxides including ZnO\cite{zhu2012observation}, HfO$_2$\cite{Xavier2013}, TiO$_2$\cite{hu2013high}, Ta$_2$O$_5$\cite{chen2013conductance} and V$_2$O$_5$\cite{yun1993room}. Specifically, during the `set' (switching from the off-state to the on-state) and/or the opposite `reset' process, the measured conductance exhibits a sequence of states with a distribution of peaks at integer multiples (from one to a few) of the conductance quantum $G_0=2e^2/h$.
Understanding the resistive switching phenomena is crucial towards reducing device variance, which is considered as a key challenge to bring memristive switching devices from lab to market\cite{yang2013memristive}. Observations of the evolution of the conductance during set or reset processes, including conductance quantization as well as subsequent pinch-off of the conductance during reset, provide ideal opportunities to uncover the switching mechanism.  %In this work, we use density functional theory (DFT) to model HfO$_2$-based devices. We specifically choose HfO$_2$ because %as the oxide as this material is particularly interesting because 
%of its compatibility %of HfO$_2$ 
%with complementary
%metal oxide semiconductor (CMOS) processing and high scalability
%\cite{de2012semiconducting,sharath2014towards,sharath2014thickness,zhang2014set,lin2011electrode,matveyev2015resistive,lian2014three,nakamura2016competitive}. 
%
It is believed that resistive switching in HfO$_2$-based devices is an atomistic process involving motion of oxygen in the oxide towards the positive electrode, which creates a reduced oxide with oxygen vacancies\cite{yang2013memristive,Francesca2013}. %DFT calculations can serve as a powerful tool to investigate atomic-scale processes in devices. 
Based on DFT calculations, a linear filament model of oxygen vacancy (V$_o$) has been proposed\cite{lian2014multi,Xavier2012} for HfO$_2$-based devices. This filament model predicts quantized conductance consistent with experimental work\cite{Xavier2013}, wherein quantized conductance is observed in Pt/HfO$_2$/Pt structures in the last few on-states before reaching an insulating off-state. However, knowledge of how the system switches between on-states and off-states is still lacking. %, which impedes understanding of the processes in working devices. 
In this work we use DFT-based quantum transport simulations 
(see Supplemental Material for details about the computational methods) to model the conductance evolution of a HfO$_2$-based device when one or two oxygen atoms move within an oxygen-vacancy rich filament during set or reset processes. %, and address the general questions of how the motion of a few (one or two) oxygen atoms along a conducting filament affects the electronic structure and conduction, and how far the oxygen atoms have to move to essentially make the filament insulating. 
Our calculations show, remarkably, that one (two) oxygen atoms effectively divides the V$_o$ filament into two (three) potential wells with midgap states, which can be well described by one-dimensional potential well states. The conductance of the device depends very sensitively on the location of the blocking oxygen atoms and can change orders of magnitudes when a single oxygen atom moves~$\alt$1~nm.

\begin{figure}[t]%[h!]
\centering
\includegraphics[width=0.45\textwidth]{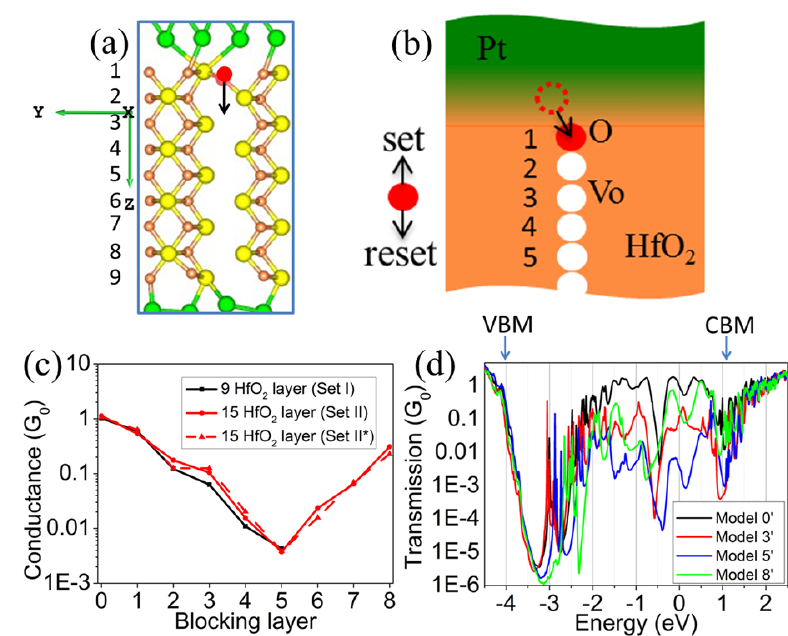}
\caption{(a): optimized atomic structure of Model 1 in device Set \textbf{I} (see text). Green, yellow, and pink colors represent Pt, Hf, and O atoms, respectively, with the blocking oxygen highlighted in red. The numbers in both (a) and (b) represent oxide atomic layers in the HfO$_2$; (b): cartoon of the modeled reset process in which a single oxygen atom moves from the top Pt electrode to block the V$_o$ filament in the oxide film; (c): evolution of the device conductance in units of $G_0$ for both Set \textbf{I} and Set \textbf{II} as an oxygen atom blocks the V$_o$ filament at different atomic layers. The dashed line (Set II*) uses a smaller energy range to estimate the conductance (see text); (d): electron transmission ($T$) in units of $G_0$ as a function of energy for selected models in Set \textbf{II}. The locations of the conduction band minimum (CBM) and the valence band maximum (VBM) of stoichiometric HfO$_2$ are indicated at the top of the panel. It is clear that transmission around the device Fermi level (set to zero energy) is dominated by the V$_o$ filament.} 
\label{fig:combine}
\end{figure}

We have modeled two sets,  Set~\textbf{I} and 
Set~\textbf{II}, of a Pt-HfO$_2$-Pt device to include the effect of the switching material thickness on the electronic properties. Set \textbf{I} (Set \textbf{II}) is based on a structure with nine (15) cubic HfO$_2$ atomic layers with the total thickness of 1.6~nm (2.7~nm). %, and Set \textbf{II} is based on a structure with 15 atomic layers with the total thickness of 2.7~nm. 
Six Pt atomic layers are always included on both sides of the oxide simulating the electrodes. For both sets we first consider the case of a fully open filament by removing from each oxide layer an adjacent oxygen atom (Model 0 in Set~\textbf{I} and Model 0' in Set~\textbf{II}), forming a metallic oxygen vacancy (V$_o$) filament penetrating the entire oxide film\cite{Xavier2012,C6CP00450D}. For both sets we also consider the case where the filament is blocked by a single oxygen atom. We denote different atomic models by Model 1 (1'), Model 2 (2'), and so on for Set \textbf{I} (Set \textbf{II}). These models are constructed by placing a single oxygen atom at the oxide atomic layer corresponding to the index of each model [see Fig.~\ref{fig:combine} (b) for an illustration of Model 1]. Because we adopt an approximately symmetric model\footnote{We note that the filament model is not exactly symmetric because of different oxide terminations at the two Pt interfaces. This will result in a very slight asymmetry in properties such as projected density of states near the interfaces.} with 9 (15) HfO$_2$ atomic layers in Set \textbf{I} (Set \textbf{II}), 5 (8) models are included to represent all possible independent blocked cases in Set \textbf{I} (Set \textbf{II}). We also consider representative models of two oxygen atoms blocking the filament for Set \textbf{II}, with the results for the low-bias conductance shown in Table~\ref{table:length}.  

\begin{table}[h!]
\caption{Low-bias conductance in units of $G_0$ for representative models in device Set \textbf{II} when two oxygen atoms block the conducting filament. Model 1'-2' denotes the case where one oxygen atom is located at the first oxide atomic layer and the other at the second oxide layer, and so on.}
\begin{center}
\begin{tabular}{ | c | c | c | c | c | c | }
 \hline
  Model &\hspace{0.2cm}1'-2'\hspace{0.2cm} &\hspace{0.2cm}1'-5'\hspace{0.2cm}&\hspace{0.2cm}1'-8'\hspace{0.2cm}&\hspace{0.2cm}4'-8'\hspace{0.2cm}&\hspace{0.2cm}7'-8'\hspace{0.2cm} \\
 \hline
  \hspace{0.2cm}conductance\hspace{0.4cm}&0.053\hspace{0.2cm}&\hspace{0.2cm}0.011\hspace{0.2cm}&\hspace{0.2cm}0.37\hspace{0.2cm}&\hspace{0.2cm}0.014\hspace{0.2cm}&\hspace{0.2cm}0.00076\hspace{0.2cm}\\
  \hline
\end{tabular}
\end{center}
\label{table:length}
\end{table}

Figure~\ref{fig:combine} (c) depicts how the low-bias conductance evolves when one blocking oxygen atom moves inside the V$_o$ filament. The low-bias conductance is estimated as the averaged electron transmission in an energy range from E$_f$-0.2~eV to E$_f$+0.2~eV (E$_f$ being the Fermi energy), corresponding to read-out voltages in the range of 0.4~V, which is typical for memristors. We have also evaluated the conductance in a smaller energy range from E$_f$-0.1~eV to E$_f$+0.1~eV, corresponding to a read-out voltage in the range of 0.2~V [Figure~\ref{fig:combine} (c), dashed line], which shows the same trend of conductance variation. For both Set \textbf{I} and Set \textbf{II}, the device conductance decreases exponentially when one blocking oxygen moves from the interface to the fifth oxide layer. Figure~\ref{fig:combine} also shows that for Set \textbf{II} the conductance rather unexpectedly increases exponentially as the oxygen atoms moves from layer 5 to layer 8, which is at the midpoint of the V$_o$ filament. The electron transmission probability\cite{Ivan2008, rocha2005, rocha2006}, $T$, underlying the conductance, is also plotted in Fig.~\ref{fig:combine} (d) as a function of energy for a single blocking oxygen atom. This figure shows that $T$ at E$_f$, and also in the whole spectrum where the conductance is dominated by vacancy states (from E$_f$-3~eV to E$_f$+1~eV), is rapidly reduced as the blocking oxygen atom moves towards the fifth layer from interface, and then enhanced as the oxygen moves from the fifth layer to the eighth layer. Table~\ref{table:length} shows that the conductance of a filament with two blocking oxygen atoms varies greatly (by three orders of magnitude) depending on the locations of the two atoms.

In the experimental work of Ref.~\onlinecite{Xavier2013}, the low-resistance states have a large conductance of about $10^2$~$G_0$, signaling conducting filaments with a large diameter and many conducting channels. During the reset process the conductance is at first reduced to one or a few $G_0$, which shows that many of the channels are pinched off and the device is entering a one-dimensional quantum regime. The transition from these intermediate conducting states to the final high-resistance states is very abrupt, with the final conductance attaining a value of about $10^{-2}$~$G_0$. Our calculated conductance for the fully open filament is 1.02 and 1.13~$G_0$ for Model 0 and 0', respectively. This is consistent with the experimental distribution of the conductance of the intermediate states, which has a peak at $G_0$. When an oxygen atom moves into the filament, the conductance decreases exponentially, and with the oxygen at the fourth oxide layer, the conductance becomes two orders of magnitude smaller than that of the open V$_o$ filament. Because the interlayer spacing between HfO$_2$ atomic layers is 1.8~{\AA},
%~{\AA}. 
the Pt-HfO$_2$-Pt device is in this case effectively switched off by a single oxygen atom moving less than one~nm. When a second oxygen atom moves into the filament, the conductance is in general further constrained (Table~\ref{table:length}), and in most cases the conductance ranges from $10^{-3}$~$G_0$ to $10^{-2}$~$G_0$. Therefore, the modeled filament with one or two blocking oxygen atoms can account for the experimentally observed very large resistance variation in the high-resistance states, which spans a range from $10^{4}$~$\Omega$ to more than $10^{7}$~$\Omega$ (or equivalently, from about $10^{-3}$~$G_0$ to $G_0$ in conductance).  

\begin{figure}[t]%[h!]
\centering
\includegraphics[width=0.47\textwidth]{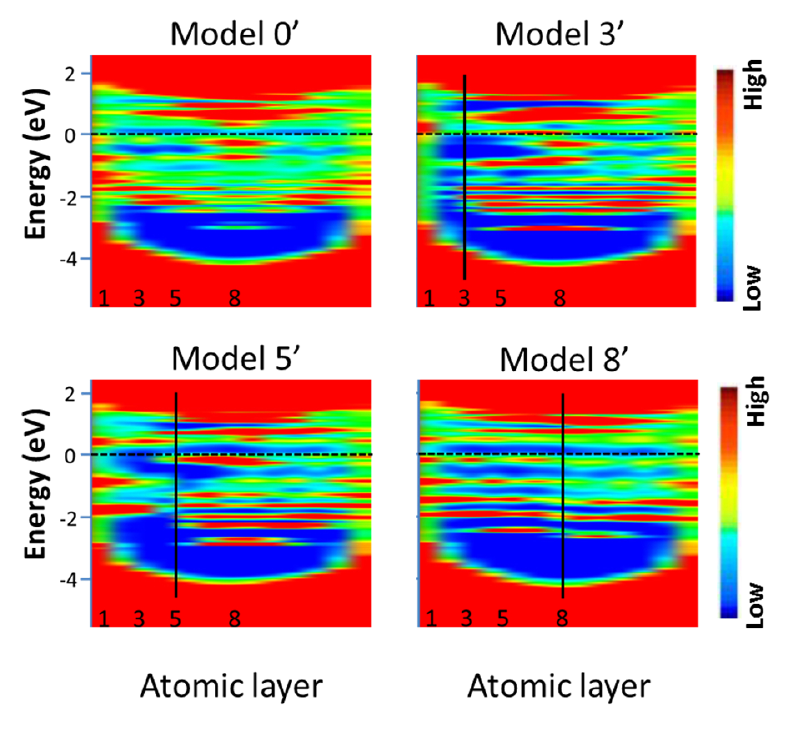}
\caption{Local density of states (LDOS) of the HfO$_2$ film sandwiched between Pt electrodes (selected models in Set \textbf{II}). The horizontal axis represents the $z$-axis with the numbers 1, 3, 5,...8, denoting the location of the corresponding oxide atomic layer. The vertical solid line shows the position of the blocking oxygen atom. The color maps denote the LDOS in arbitrary units. The Fermi level is set at zero energy and indicated by dotted horizontal lines.} 
\label{fig:QW}
\end{figure}

\begin{figure}[t]%[h!]
\centering
\includegraphics[width=0.47\textwidth]{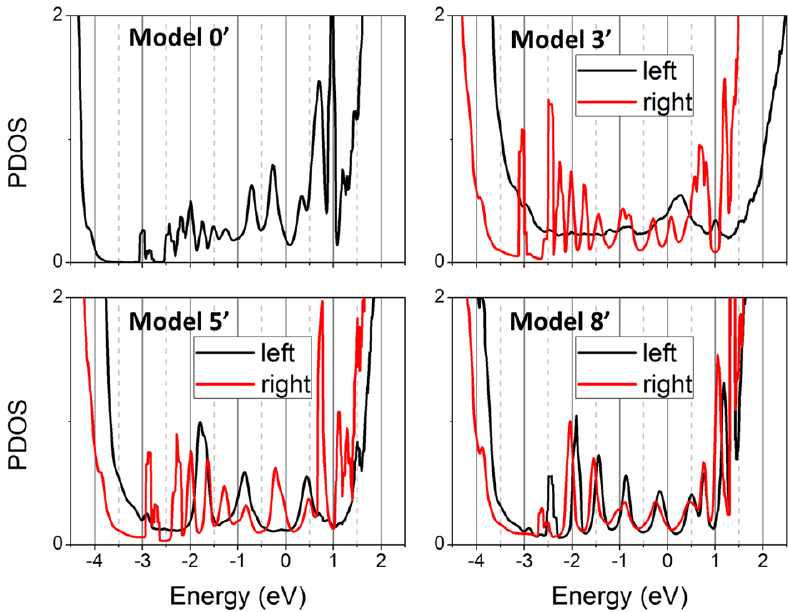}
\caption{Averaged PDOS of QWs (see text) for selected models. 'left' ('right') denote the averaged PDOS to the left (right) of the blocking atom. The Fermi level is set at zero energy.} 
\label{fig:pdos}
\end{figure}

In order to understand the predicted features in $T$ we now analyze the electronic structures of the modeled devices. We will focus on Set \textbf{II}
as the same switching mechanism applies to both sets (Supplemental Material). Figure~\ref{fig:QW} shows the local density of states (LDOS) of the sandwiched oxide for selected models of Set \textbf{II}. The blue regions indicate negligible LDOS in the fundamental energy gap of HfO$_2$. The outermost oxide atomic layers exhibit finite LDOS in the full energy range as a result of hybridization of the Pt-HfO$_2$ states at the interface, giving rise to midgap evanescent interface states. The oxide LDOS also shows a clear band bending up to the 5th atomic layer from the Pt interface. Within the blue midgap regions there exist bright stripes, which are the midgap eigenstates in the finite length V$_o$ filament. We note that these states have the appearance of one-dimensional (1D) quantum well (QW) states. For model 0' there is no oxygen atom blocking the filament, and these midgap eigenstates are confined by the two Pt/HfO$_2$ interfaces on both sides. The blocking oxygen atom divides the V$_o$ filament into two QWs for models 3', 5' and 8' (Fig.~\ref{fig:QW}). A similar analysis has been used previously to compare modeling results to experimental data to show that hydrogen atoms behave as semitransparent barriers when binding to silicon nanowires\cite{naydenov2015single}.  Fig.~\ref{fig:QW} also shows a strong dependence of number of QW states on QW length, with more midgap states for longer QWs.

To help analyze the nature of the midgap states and to correlate the electronic structure with transport properties, the averaged projected density of states (PDOS) is shown in Fig.~\ref{fig:pdos}. The average PDOS is obtained by calculating the LDOS on the left and right side of the blocking oxygen atom, respectively, and averaging over the thickness of the left and right side. Comparing Figs.~\ref{fig:QW} and \ref{fig:pdos} we note that each PDOS peak corresponds to a QW eigenstate (a bright stripe in Fig.~\ref{fig:QW}). The number of QW states in the midgap region (from about -3~eV to 1~eV) is nearly linear with the QW length, which is consistent with a 1D square well model. Fits to a square well with infinitely high barriers using the free electron mass and using the well length as a fitting parameter gives a fitted well length typically a factor two larger than the actual length. This is not unreasonable, given the deviations of the actual carrier mass from the ones of free electrons, as well as of the actual boundary conditions at the edges of each well, which are probably much softer than those of an infinite potential well.

In order to elucidate the switching mechanism, we now correlate the electronic structure (PDOS) with transmission. In bulk HfO$_{2-x}$, the oxygen vacancies give rise to a small DOS in the gap region because of two vacancy-induced bands\cite{C6CP00450D}. When the oxide with a conducting filament without a blocking oxygen atom is sandwiched between two electrodes, there exist discrete vacancy-induced QW states instead of a dispersive band. This is seen as a number of peaks in PDOS in the midgap region (Fig.~\ref{fig:pdos} Model 0'); at other energies in the gap states are evanescent from the electrode and transmission only occurs through tunneling. However, at energies in resonance with the QW states, transmission occurs through resonant tunneling, which gives rise to peaks in $T$ (Fig.~\ref{fig:combine} d). Because of the relatively long QW and the intrinsic broadening of the QW states, PDOS is finite and pseudo-metallic in a large range of the gap region (from $\sim-3$~eV and up), and $T$ is large in this range of energy. Because of a relatively large PDOS at the Fermi level, due to two large peaks at about -0.25~eV and $+0.25$~eV, the low-bias conductance is also large. 

When the oxygen atom is introduced into the filament, it divides the filament into two QWs and acts as a semi-transparent barrier between the two QWs. The transmission probability is then the result of superimposing tunneling with resonances in the two QWs. With the oxygen atom in any of the first few layers, the PDOS on the left side of the oxygen (the short  QW) is large and finite across the energy gap because all evanescent states in this energy range have a finite probability density in this short QW. To the right of the oxygen atom, the QW is long and there are a number of states, clearly visible in PDOS (Fig.~\ref{fig:pdos} b). As the oxygen atom moves from the interface, the tunneling probability and PDOS to the left of the oxygen atom decay exponentially. This causes the overall exponential decay of $T$ in the whole energy range of the energy gap, from about $-3$~eV to 1~V, even though there are clear peaks that arise from accidental degeneracies in the asymmetric QWs. This mechanism leads to the low-bias conductance, which is determined by $T$ around $E_f$, to decrease exponentially (Fig.~\ref{fig:combine}). Energy calculations show that the total energy of the device is lowest when the blocking oxygen atom is at the center
of the filament, consistent with a weak coupling between degenerate QW states (see Supplemental Material).
%with some peaks due to accidental degeneracies in the two QWs.
As the oxygen atom moves from layer 5 (Model 5') towards the center (Model 8'), the QWs now become more symmetric (Fig.~\ref{fig:pdos} lower right panel), and the eigenstates in the two QWs become degenerate. The resonant peaks in $T$ then grow rapidly because of resonant transport. In particular, $T$ increases rapidly near $E_f$. As a consequence, the low-bias conductance increases quickly as the blocking oxygen atom moves from position 5 to the center (position 8). This growth gives rise to the rapid \textit{increase} in the low-bias conductance (Fig.~\ref{fig:combine} (c)). We have also performed one calculation for a 19 oxide layers long filament and with a blocking atom at the center in order to see if the conductance saturates with device length. The low-bias conductance is calculated to be 0.35~$G_0$, which is quite similar to the value of 0.31~$G_0$ for device Set \textbf{II}. This indicates that the conductance with one oxygen atom at the $V_o$ filament center indeed saturates as the filament length increases beyond that of Set \textbf{II} (15 atomic layers). We note a similar analysis applies to the shorter device, Set \textbf{I}.

\begin{figure}[t]%[h!]
\centering
\includegraphics[width=0.47\textwidth]{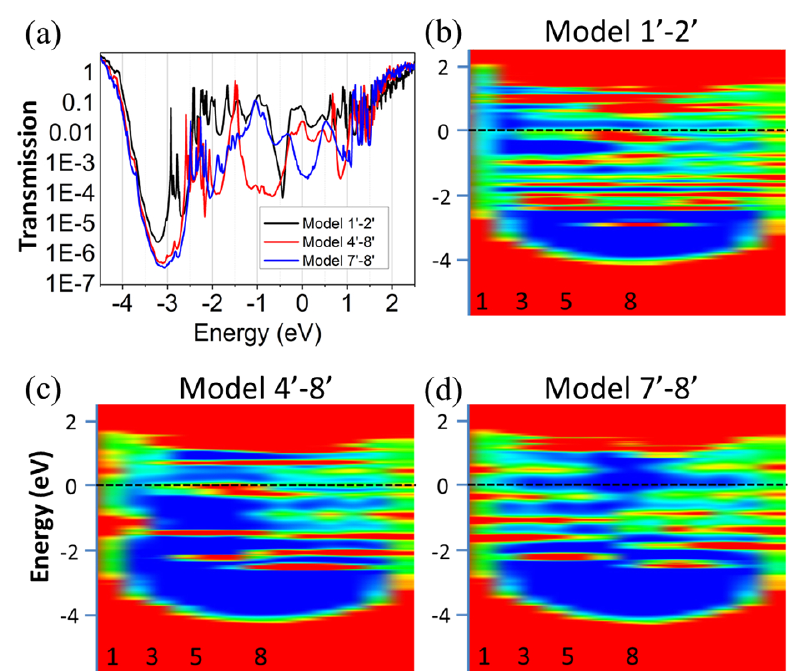}
\caption{(a): Transmission in units of $G_0$ as a function of energy for representative models in the case of two blocking oxygen atoms. (b) to (d): corresponding LDOS of the two-blocking-atom models in arbitrary unit. For all plots device Fermi level is aligned to zero energy.} \label{fig:combine2}
\end{figure}

When two oxygen atoms block the filament simultaneously device conductance is in general further suppressed (Table 1). On the other hand, there appears to be strong dependence of transmission on the locations of both atoms (Fig.~\ref{fig:combine2}). Transmission of Model 1'-2' is in general higher in the midgap range than that of Model 4'-8' or Model 7'-8', since in Model 1'-2' there is only one QW (Fig.~\ref{fig:combine2} (b)) to the right of the two blocking atoms, and eigenstates in this QW can couple with the left electrode by evanescent interface states. The comparison between Model 4'-8' and Model 7'-8' is a little subtle. In general, transmission peaks due to resonance of QW states have similar heights for both models. The difference in low-bias conductance of about one order of magnitude (Table 1) origins from the detailed distribution of QW states in energy. For Model 4'-8' which has three QWs there are eigenstates around E$_f$ in both the middle and the right QWs (Fig.~\ref{fig:combine2} (c)). These states can couple with the left electrode by interface states, resulting in a transmission peak around E$_f$. For Model 7'-8' there are two QWs separated by two adjacent blocking atoms. On each side of the blocking atoms E$_f$ locates at about the middle of two midgap states, resulting in a suppression of transmission about E$_f$ (Fig.~\ref{fig:combine2} (a)). 

\begin{acknowledgments}
The work by X.Z., P.Z. and O.H. was supported by U. S. DOE, Office of Science under Contract No. DE-AC02-06CH11357. I.R. acknowledges financial support from the European Union's Horizon2020 research and innovation programme within the PETMEM project (Grant agreement number 688282). We gratefully acknowledge the computing resources provided on Blues and Fusion, high-performance computing clusters operated by the Laboratory Computing Resource Center at Argonne National Laboratory. 
\end{acknowledgments}

%\section{METHODS}

\bibliographystyle{aipnum4-1}
\bibliography{sample}

\end{document}